\documentclass[10pt,conference]{IEEEtran}
\usepackage{times}
\def\beginofproof{\noindent {\bf Proof. }}
\def\endofproof{\hfill\rule{6pt}{6pt}}
\newtheorem{theorem}{Theorem}
\newtheorem{lemma}{Lemma}

\newtheorem{corollary}{Corollary}
\newtheorem{example}{Example}

\begin{document}

\title{Complete Enumeration of Stopping Sets of Full-Rank Parity-Check
Matrices of Hamming Codes}

\author{\authorblockN{Khaled A.S.\ Abdel-Ghaffar}
\authorblockA{University of California \\
Department of ECE \\
Davis, CA 95616 \\
USA \\
E-mail: ghaffar@ece.ucdavis.edu} \and
\authorblockN{Jos H. Weber}
\authorblockA{Delft University of Technology \\
Faculty of EEMCS \\
Mekelweg 4, 2628 CD Delft \\
The Netherlands \\
E-mail: j.h.weber@ewi.tudelft.nl}
 }

\maketitle

\begin{abstract}
Stopping sets, and in particular their numbers and sizes,
play an important role in determining the performance of iterative
decoders of linear codes over binary erasure channels.
In the 2004 Shannon Lecture, McEliece presented an expression for
the number of stopping sets of size three for a full-rank
parity-check matrix of the Hamming code. In this correspondence, we derive
an expression for the number of stopping sets of any given size for the same
parity-check matrix.
\end{abstract}

\


\section{Introduction} \label{intro}

Let $\cal C$ be a linear binary $[n,k,d]$ block code, where $n$,
$k$, and $d$ denote the code's length, dimension, and Hamming
distance, respectively. The set of codewords of $\cal C$ can be
defined as the null space of the row space of an $r\times n$
binary parity-check matrix ${\bf H}=(h_{i,j})$ of rank $n-k$.
Assuming all rows in $\bf H$ are different, $n-k\le r\le 2^{n-k}$.

Let ${\cal S}$ be a subset of $\{1,2,\ldots,n\}$ and
${\cal T}$ be a subset of $\{1,2,\ldots,r\}$.
For any ${\bf H}=(h_{i,j})$ of size $r\times n$,
let ${\bf H}_{\cal S}^{\cal T}=(h_{i,j})$ where $i\in{\cal T}$ and
$j\in{\cal S}$. Then, ${\bf H}_{\cal S}^{\cal T}$ is a
$|{\cal T}|\times|{\cal S}|$
submatrix of ${\bf H}$. For simplicity, we write ${\bf H}_{\cal S}$
and ${\bf H}^{\cal T}$ to denote ${\bf H}_{\cal S}^{\cal T}$ in case
${\cal T}=\{1,2,\ldots,r\}$ and ${\cal S}=\{1,2,\ldots,n\}$, respectively.

A set $\cal S$ is the support of a codeword if and
only if all rows in ${\bf H}_{\cal S}$ have even weight, i.e., if
and only if
\begin{equation} \label{eqCW}
|\{j\in {\cal S}: h_{i,j}=1\}|\equiv 0 (2) \hspace{1cm}\forall
i=1,2,\ldots,r.
\end{equation}
A set $\cal S$ is a stopping set if and only if ${\bf H}_{\cal S}$
does not contain a row of weight one, i.e., if and only if
\begin{equation} \label{eqSS}
|\{j\in {\cal S}: h_{i,j}=1\}|\neq 1 \hspace{1cm}\forall
i=1,2,\ldots,r.
\end{equation}

The polynomial $A(x)=\sum_{l=0}^n A_l x^l$, where $A_l$ is the
number of codewords of weight $l$, is called the {\em weight
enumerator} of code $\cal C$. It holds that
$$d  =  \min\{l\ge 1:A_l> 0\}.$$
The polynomial $S(x)=\sum_{l=0}^n S_l x^l$, where $S_l$ is the
number of stopping sets of size $l$, is called the {\em stopping
set enumerator} of parity-check matrix $\bf H$. Let $s$ denote the
smallest size of a non-empty stopping set, i.e.,
$$s  =  \min\{l\ge 1:S_l> 0\}.$$

Notice from (\ref{eqCW}) and (\ref{eqSS}) that the support of any
codeword is a stopping set. Therefore, $S_l\ge A_l$ for $l=0,1,\ldots,n$
and $s\le d$.

Considering the vacuous case in which $l=0$, we notice, from
the definitions, that the empty set is both the support of
a codeword and a stopping set for any code and any parity-check matrix.
Hence, $A_0=S_0=1$. Furthermore, from the
observation that (\ref{eqCW}) and (\ref{eqSS}) are equivalent for
sets $\cal S$ with $|{\cal S}|\le 2$, it follows that $A_l=S_l$
for $l\le 2$. In particular, $S_0=1$ and  $S_1=S_2=0$ for any
parity-check matrix of a code of minimum distance $d\ge 3$.
On the other hand, if every row in $\bf H$ has weight at least equal
to $n-l+2$ and $\cal S$ is a subset of $\{1,2,\ldots,n\}$
of size $l$, then every row in ${\bf H}_{\cal S}$ has weight at least
equal to two. Hence, $\cal S$ is a stopping set of $\bf H$ and
\begin{equation} \label{Eq_lbig1}
S_l=\left({n \atop l}\right).
\end{equation}

The notion of stopping sets is important in the context of
iterative decoders (using $\bf H$-based Tanner graphs), in
particular for low-density parity-check codes \cite{DPTRU}. For
example, on the binary erasure channel, an iterative decoder will
not lead to successful decoding if and only if the set of erased
positions contains a non-empty stopping set. Hence, the minimum
(non-empty) stopping set size $s$ and the cardinality $S_s$ are
performance indicators for iterative decoding, alike the minimum
distance $d$ and the number of minimum weight codewords for
maximum-likelihood decoding. However, contrary to the weight
enumerator, which is fixed for a code $\cal C$, the stopping set
enumerator depends on the choice of the parity-check matrix $\bf
H$.

In the 2004 Shannon lecture, McEliece \cite{McE} presented
the following expression for the number of stopping sets
of size three in a $[2^m-1,2^m-m-1,3]$ Hamming code:
\begin{equation} \label{eqME}
S_3={1\over 6}\left(5^m-3^{m+1}+2^{m+1}\right) \sim {1\over
6}n^{2.322}.
\end{equation}
He did not mention explicitly which parity-check matrix he had in
mind, but from the context, it was clear
that it was the full rank $m\times (2^m-1)$ parity-check matrix.
This is the parity-check matrix of minimum number of rows for the
Hamming code.
The value of $S_3$ is of particular interest since $S_0=1$ and
$S_1=S_2=0$ as the Hamming code has minimum distance three
which implies that $s=3$. Hence, the value of $S_3$
given in (\ref{eqME}) corresponding to the full rank parity-check
matrix of the Hamming code can be used to obtain a good estimate
of the performance of the simplest
iterative decoder for the Hamming code. However, as the multiplicities
$A_l$ of codewords in a code of weight $l>d$
influence the performance of maximum-likelihood decoding,
the multiplicities $S_l$ of stopping sets of size $l>s$ also
influence the performance of iterative decoding.
In this correspondence, we derive the values $S_l$ for all $l$ for
the full-rank parity-check matrix of the Hamming code.

Section~II gives an expression for $S_l$ in terms of certain
parameters that can be derived from the parity-check matrix of
any linear code. In Section~III, we determine these parameters for the full-rank
parity check matrices of Hamming codes and derive expressions for their
stopping set enumerators.

\section{Computing the Stopping Set Enumerator for Linear Codes}
In this section, we derive an expression for the coefficients of
$S(x)$ for an $r\times n$ parity-check matrix $\bf H$ of an $[n,k,d]$
binary linear code $\cal C$.

For each subset ${\cal T}$ of $\{1,2,\ldots, r\}$, we define
$z_{\cal T}$ to be the number of columns in ${\bf H}^{\cal T}$ of
weight zero. In particular, if $\cal T$ is empty, then $z_{\cal
T}=n$. If ${\cal Y}$ is a subset of $\{1,2,\ldots,n\}$ such that
$|{\cal Y}|=p$, each row in ${\bf H}_{\cal Y}^{\cal T}$ has weight
one, and each column in ${\bf H}_{\cal Y}^{\cal T}$ has non-zero
weight, then we say that ${\cal Y}$ is of type $p$ with respect to
${\cal T}$. Let $Y({\cal T},p)$ be the number of subsets of
$\{1,2,\ldots,n\}$ of type $p$ with respect to ${\cal T}$. Since
every column in ${\bf H}_{\cal Y}^{\cal T}$ should have non-zero
weight and every row should have weight one, it follows that
\begin{equation} \label{Y0}
Y({\cal T},p)=\left\{\begin{array}{ll} 0 & \mbox{if ${\cal
T}=\emptyset\wedge p\ge 1$ or ${\cal T}\ne\emptyset\wedge p=0$},
\\ 1 & \mbox{if ${\cal T}=\emptyset\wedge p=0$}.\end{array}\right.
\end{equation}

\begin{example}
Consider the full-rank parity-check matrix $\bf H$ of the $[7,4,3]$ Hamming
code given by
$${\bf H}=\left[\begin{array}{ccccccc}
1 & 0 & 1 & 0 & 1 & 0 & 1\\
1 & 1 & 0 & 0 & 1 & 1 & 0\\
1 & 1 & 1 & 1 & 0 & 0 & 0\end{array}\right].$$
Let ${\cal T}=\{2,3\}$. Then, $z_{\cal T}=1$ as ${\bf H}^{\cal T}$ has exactly
one all-zero column and $Y({\cal T},2)=4$ since there are exactly four subsets
of type $p=2$ with respect to ${\cal T}=\{2,3\}$ given by $\{3,5\}$, $\{3,6\}$,
$\{4,5\}$, and $\{4,6\}$. Indeed,
${\bf H}_{\cal Y}^{\cal T}$ is a $p\times |{\cal T}|$ matrix each row
of which has weight one and each column of which is nonzero if and only if
$\cal Y$ is one of these four subsets.
\end{example}

\begin{theorem} \label{Thm1}
Let $\bf H$ be an $r\times n$ parity-check matrix of a linear
code $\cal C$ of length $n$.
Then, for $l=0,1,\ldots,n$,
$$S_l=\sum_{{\cal T}\subseteq\{1,2,\ldots,r\}}(-1)^{|{\cal T}|}\sum_{p=0}^l
                    Y({\cal T},p)\left({z_{\cal T}\atop l-p}\right).$$
\end{theorem}
\beginofproof It follows from (\ref{Y0}) that the result holds for $l=0$.
Therefore, we assume $l\ge 1$.
Let $\cal S$ be a non-empty subset of $\{1,2,\ldots,n\}$.
Following \cite{SV1}, we say that row $\bf h$ in $\bf H$ covers $\cal S$
if ${\bf h}_{\cal S}$ has weight one. For $i=1,2,\ldots,r$ and
$l=1,2,\ldots, n$, let ${\cal Q}_i^l$ and $\overline{{\cal Q}_i^l}$ be the family
of subsets of $\{1,2,\ldots, n\}$ and size $l$ that are covered and not
covered, respectively, by the $i$th row in $\bf H$.
A subset ${\cal S}\subseteq\{1,2,\ldots,n\}$ of size $l$ is a stopping
set if and only if it is not covered by any row of $\bf H$, i.e.,
${\cal S}\in\cap_{i=1}^r \overline{{\cal Q}_i^l}$.
In particular, $S_l=|\cap_{i=1}^r \overline{{\cal Q}_i^l}|$.
The Principle of Inclusion and Exclusion (see, e.g., \cite{COM74}, Theorem B, p. 178), yields
\begin{equation} \label{Eq_PIE}
|\cap_{i=1}^r \overline{{\cal Q}_i^l}|=\sum_{{\cal T}\subseteq\{1,2,\ldots,r\}}
                      (-1)^{|{\cal T}|}|\cap_{i\in{\cal T}}{\cal Q}_i^l|.
\end{equation}
In the above sum, ${\cal T}$ runs over all $2^r$ subsets of $\{1,2,\ldots,r\}$
where, for the term corresponding to the empty set $\emptyset$,
$$|\cap_{i\in\emptyset}{\cal Q}_i^l|=\left({n\atop l}\right),$$
which is the total number of subsets of $\{1,2,\ldots,n\}$ of size $l$.
Notice that a set ${\cal S}$ of size $l$ belongs to
$\cap_{i\in{\cal T}}{\cal Q}_i^l$ if and only if
each row in ${\bf H}_{\cal S}^{\cal T}$ has weight one. This is the case
if and only if, for some $p=0,1,\ldots,l$, ${\cal S}$ contains a subset $\cal Y$
of type $p$ with respect to ${\cal T}$ and all the columns indexed by
${\cal S}\backslash{\cal Y}$ have zero weights. Therefore,
$$|\cap_{i\in{\cal T}}{\cal Q}_i^l|=\sum_{p=0}^l
                   Y({\cal T},p)\left({z_{\cal T}\atop l-p}\right).$$
Combining this with (\ref{Eq_PIE}), the proof is complete.
\endofproof

\section{Computing the Stopping Set Enumerator for Hamming Codes}
A full-rank parity-check matrix of a $[2^m-1,2^m-m-1,3]$ Hamming code, where $m\ge 3$,
is an $m\times (2^m-1)$ matrix whose columns are the distinct non-zero vectors
of length $m$. From now on, we take ${\bf H}$ to be such a matrix with
$r=m$ and $n=2^m-1$, and $S_l$ to be the number of stopping sets of $\bf H$ of size $l$.
In particular, since every row in $\bf H$ has weight $2^{m-1}$, we have from (\ref{Eq_lbig1})
$$S_l=\left({2^m-1\atop l}\right)$$
for $2^{m-1}+1\le l\le 2^m-1$. 
Our goal is to determine $S_l$ for all $l$. 

In the following derivations, we make use of Stirling numbers.
Following the notation of Comtet \cite{COM74} and Riordan \cite{Riordan78},
we denote by $s(n,k)$ and $S(n,k)$
the Stirling numbers of the first kind and of the second kind, respectively.
Notice that $n$ and $k$ are not necessarily the length and the dimension of a code.
For $n\ge 1$, $(-1)^{n-k}s(n,k)$
is the number of permutations of $n$ elements which have exactly $k$ cycles and
$S(n,k)$ is the number of ways of partitioning a set of $n$ elements into $k$ non-empty
subsets, see, e.g., \cite{AS64}, p. 824, where (\ref{Eq_s1})--(\ref{Eq_Sexp}) can also be found.
We also define $s(0,0)=S(0,0)=1$. Notice that for $n\ge 1$, both $s(n,k)$ and $S(n,k)$ are
equal to zero if $k\le 0$ or $k>n$, see e.g., \cite{Riordan78}.

Stirling numbers of the first kind satisfy the following polynomial identity in $x$
\begin{equation} \label{Eq_s1}
\sum_{k=0}^n s(n,k)x^k=n!\left({x \atop n}\right)
\end{equation}
and the recursion
\begin{equation} \label{Eq_Rs}
s(n+1,k)=s(n,k-1)-ns(n,k),
\end{equation}
for $n\ge k\ge 1$.
We also have
\begin{equation} \label{Eq_s2}
s(n,1)=(-1)^{n-1}(n-1)!.
\end{equation}

Stirling numbers of the second kind can be computed explicitly using
\begin{equation} \label{Eq_Sexp}
S(n,k)={1\over k!}\sum_{i=0}^k (-1)^i\left({k\atop i}\right)(k-i)^n.
\end{equation}

The reader can find a lucid treatment of Stirling numbers of both kinds in
\cite{GKP89} but with different notation and where Stirling numbers of the first
kind are defined as $(-1)^{n-k}s(n,k)$.

\begin{lemma}  \label{Lem_bins}
For $n\ge 0$, the following polynomial identity in $x$ holds
$$n! \left({x-1\atop n}\right)=\sum_{k=1}^{n+1}s(n+1,k)x^{k-1}.$$
\end{lemma}
\beginofproof From the definition of binomial coefficients, we have
$$\left({x-1\atop n}\right)={x-n\over x}\left({x\atop n}\right).$$
Combining this with (\ref{Eq_s1}) we get
\begin{eqnarray*}
n! \left({x-1\atop n}\right)&=& {x-n\over x}\sum_{k=0}^n s(n,k) x^k\\
                            &=&\sum_{k=0}^n s(n,k)x^k-n\sum_{k=0}^n s(n,k)x^{k-1}\\
                            &=&\sum_{k=1}^{n+1} s(n,k-1)x^{k-1}-n\sum_{k=0}^n s(n,k)x^{k-1}\\
                            &=&\sum_{k=1}^{n+1}\left(s(n,k-1)-ns(n,k)\right)x^{k-1}
\end{eqnarray*}
since $ns(n,0)=0$ and $s(n,n+1)=0$ for $n\ge 0$. The result now follows from the
recurrence relation (\ref{Eq_Rs}).
\endofproof

\begin{lemma} \label{Lem_Ham1}
Let $\bf H$ be an $m\times (2^m-1)$ parity-check matrix of a Hamming code of
length $2^m-1$.
Then, for any set ${\cal T}\subseteq\{1,2,\ldots,m\}$ and any $p=0,1,\ldots,n$,
$$ z_{\cal T}=2^{m-|{\cal T}|}-1$$
and
$$ Y({\cal T},p)=S(|{\cal T}|,p)\;2^{(m-|{\cal T}|)p}. $$
\end{lemma}
\beginofproof Each non-zero vector of
length $m$ appears exactly once as a column in $\bf H$. Hence,
every zero-weight vector of length $|{\cal T}|$ appears exactly
$2^{m-|{\cal T}|}-1$ times in ${\bf H}^{\cal T}$. This verifies
the expression for $z_{\cal T}$. From (\ref{Y0}) and
(\ref{Eq_Sexp}), it follows that the second part of the lemma holds in
case ${\cal T}=\emptyset$. Therefore, assume in the following that
${\cal T}\ne\emptyset$. Hence, $1\le|{\cal T}|\le m$. For each
$j\in \{1,2,\ldots,2^m-1\}$, let
$${\cal H}^{\cal T}_j=\{i\in {\cal T}: h_{ij}=1\},$$
i.e., ${\cal H}^{\cal T}_j$ is the set of indices of ones in the $j$th column
of ${\bf H}^{\cal T}$.
Notice that a subset ${\cal Y}$ of size $p$ is of type $p$ with respect to $\cal T$
if and only if ${\bf H}_{{\cal Y}}^{{\cal T}}$ is a $|{\cal T}|\times p$ matrix,
each column of which has non-zero weight, and
each row of which has weight one. This is the case if and only if, for $j\in{\cal Y}$,
the sets ${\cal H}^{\cal T}_j$ are non-empty, disjoint, and their union is $\cal T$.
There are $S(|{\cal T}|,p)$ ways to partition a set of $|{\cal T}|$ elements into $p$
non-empty disjoint subsets. Each non-zero
vector of length $|{\cal T}|$ appears exactly $2^{m-|{\cal T}|}$ times as a column
in ${\bf H}^{\cal T}$. Therefore, for each partition, there are
$2^{m-|{\cal T}|}$ choices for $j$,
corresponding to indices of identical non-zero columns in ${\bf H}^{{\cal T}}$, such that
${\cal H}^{\cal T}_j$ is one of the $p$ subsets in the partition.
Since there are $p$
subsets in each partition, and there are $S(|{\cal T}|,p)$ partitions, the total
number of $|{\cal T}|\times p$ submatrices in ${\bf H}^{{\cal T}}$,
where $0\le p\le l$,
each of its
columns has non-zero weight and each of its rows has weight one, is
$S(|{\cal T}|,p)\;2^{(m-|{\cal T}|)p}$.
The number of such matrices equals $Y({\cal T},p)$.
\endofproof

\begin{lemma} \label{Lem_Ham2}
Let $\bf H$ be an $m\times (2^m-1)$ parity-check matrix of a Hamming code of
length $2^m-1$.
Then, for $l=0,1,\ldots,2^m-1$,
$$S_l=\sum_{t=0}^m (-1)^t\left({m\atop t}\right)\sum_{p=0}^l S(t,p) 2^{(m-t)p}
                    \left({2^{m-t}-1\atop l-p}\right).$$
\end{lemma}
\beginofproof From Lemma~\ref{Lem_Ham1}, it follows that for fixed $m$ and $p$,
both $z_{\cal T}$ and $Y({\cal T},p)$ depend only on the cardinality of $\cal T$.
Since there are ${m \choose t}$ subsets $\cal T$ of cardinality $t$, then the
result follows by using the expressions derived for $z_{\cal T}$ and $Y({\cal T},p)$
in Theorem~\ref{Thm1}.
\endofproof

Lemma~\ref{Lem_Ham2}, when combined with (\ref{Eq_Sexp}), gives an explicit expression
for $S_l$ for $l=0,1,\ldots,2^m-1$.
Notice that for each value of $l$, $S_l$ depends only on $m$.
However, contrary to the expression in (\ref{eqME}),
the expression in the lemma does not show clearly this dependency.
In the next lemma, we make the dependency clear.
For this purpose, we define for nonnegative integers $q$ and $v$,
\begin{equation} \label{Eq_b}
b(q,v)=\sum_{p=0}^{v}(-1)^p \left({v\atop p}\right)
                      s(p+1,p-q+1).
\end{equation}

\begin{lemma} \label{Lem_Ham3}
Let $\bf H$ be an $m\times (2^m-1)$ parity-check matrix of a Hamming code of
length $2^m-1$.
Then, for $l=0,1,\ldots,2^m-1$,
$$S_l={1\over l!}\sum_{q=0}^l \sum_{v=0}^l (-1)^v\left({l\atop v}\right)b(q,v)(2^{l-q}-(l-v))^m.$$
\end{lemma}
\beginofproof From Lemma~\ref{Lem_bins}, we have for $p\le l$ and $t\le m$
\begin{eqnarray}
\left({2^{m-t}-1\atop l-p}\right)\mbox{\hspace{-8pt}}&=&\mbox{\hspace{-8pt}}{1\over (l-p)!}\sum_{u=1}^{l-p+1}s(l-p+1,u)2^{(m-t)(u-1)} \nonumber \\
                                 \mbox{\hspace{-8pt}}&=&\mbox{\hspace{-8pt}}{1\over (l-p)!}\sum_{q=0}^{l-p} s(l-p+1,l-q-p+1)  \nonumber\\
                                 & &\hspace{12pt}       \cdot 2^{(m-t)(l-q-p)} \nonumber \\
                                 \mbox{\hspace{-8pt}}&=&\mbox{\hspace{-8pt}}{1\over (l-p)!}\sum_{q=0}^{l} s(l-p+1,l-q-p+1)  \nonumber\\
                                 & &\hspace{12pt}       \cdot 2^{(m-t)(l-q-p)}, \label{Eq_Thm21}
\end{eqnarray}
since $s(n,k)=0$ for $k\le 0<n$.
Also, from (\ref{Eq_Sexp}), we obtain
\begin{eqnarray}
S(t,p)&=&{1\over p!}\sum_{i=0}^p (-1)^i\left({p\atop i}\right)(p-i)^t \nonumber\\
      &=&{(-1)^{l-p}\over p!}\sum_{v=l-p}^l (-1)^v\left({p\atop l-v}\right)(l-v)^t \nonumber\\
      &=&{(-1)^{l-p}\over p!}\sum_{v=0}^l (-1)^v\left({p\atop l-v}\right)(l-v)^t. \label{Eq_Thm22}
\end{eqnarray}
since
$$\left({p\atop l-v}\right)=0$$
for $v<l-p$.
Substituting (\ref{Eq_Thm21}) and (\ref{Eq_Thm22}) in the expression of $S_l$ given in Lemma~\ref{Lem_Ham2},
we get
\begin{eqnarray*}
S_l\mbox{\hspace{-8pt}}&=&\mbox{\hspace{-8pt}}\sum_{t=0}^m (-1)^t\mbox{\hspace{-1pt}} \left({m \atop t}\right)\mbox{\hspace{-1pt}}
    \sum_{p=0}^{l}{(-1)^{l-p}\over p!}\mbox{\hspace{-1pt}}\sum_{v=0}^l (-1)^v\mbox{\hspace{-1pt}}
    \left({p\atop l-v}\right)\mbox{\hspace{-1pt}}(l-v)^t  \\
   & &\mbox{\hspace{-3pt}}\cdot {2^{(m-t)p}\over (l-p)!}\sum_{q=0}^l s(l-p+1,l-q-p+1)2^{(m-t)(l-q-p)}\\
  \mbox{\hspace{-8pt}}&=&\mbox{\hspace{-8pt}}{1\over l!}
   \sum_{t=0}^m (-1)^t\left({m \atop t}\right)\sum_{p=0}^l(-1)^{l-p}
     \mbox{\hspace{-1pt}}\left({l\atop p}\right)\mbox{\hspace{-1pt}}\sum_{v=0}^l(-1)^v\left({p\atop l-v}\right)\\
   & &\mbox{\hspace{-3pt}}\cdot(l-v)^t\sum_{q=0}^l s(l-p+1,l-q-p+1)2^{(m-t)(l-q)}.
\end{eqnarray*}
Noticing that
$$\left({l\atop p}\right)\left({p \atop l-v}\right)=\left({l\atop v}\right)\left({v\atop l-p}\right),$$
we obtain by interchanging orders of summations
\begin{eqnarray*}
S_l&=&{1\over l!}\sum_{q=0}^l\sum_{v=0}^l (-1)^v\left({l\atop v}\right)
                  \sum_{p=0}^l(-1)^{l-p}\left({v\atop l-p}\right)\\
   & &\mbox{\hspace{5pt}}\cdot s(l-p+1,l-q-p+1)\\
   & &\mbox{\hspace{5pt}}\sum_{t=0}^m(-1)^t\left({m\atop t}\right) 2^{(m-t)(l-q)}(l-v)^t\\
   &=&{1\over l!}\sum_{q=0}^l\sum_{v=0}^l (-1)^v\left({l\atop v}\right)\left(2^{l-q}-(l-v)\right)^m\\
              & &\mbox{\hspace{5pt}}\sum_{p=0}^l(-1)^{l-p}\left({v\atop l-p}\right)
                             s(l-p+1,l-q-p+1).
\end{eqnarray*}
Replacing $p$ by $l-p$
and noticing that
$$\left({v\atop p}\right)=0$$
whenever $p>v$, it follows that the last sum is identical to $b(q,v)$.
\endofproof

To obtain an expression for $b(q,v)$ that does not involve Stirling numbers, we derive
a recursion for $b(q,v)$ along with boundary values
to apply the recursion.

\begin{lemma} \label{Lem_Ham4}
For nonnegative numbers $q$ and $v$, we have
$$b(q,v)=\left\{\begin{array}{rl}
0 & \mbox{ if $q>v$ or $q=0<v$,}\\
q! & \mbox{ if $q=v$,}\\
\end{array}\right.$$
and for $q\ge 1$ and $v\ge 2$,
$$b(q,v)=vb(q-1,v-1)-(v-1)b(q-1,v-2).$$
\end{lemma}
\beginofproof For $0\le p\le v<q$, we have $p-q+1\le 0$ and, therefore,
$s(p+1,p-q+1)=0$. From the definition of $b(q,v)$ given in (\ref{Eq_b}),
it follows that $b(q,v)=0$ in case $q>v$. In case $q=0$, it follows from
the definition of $b(q,v)$ and the fact that $s(p+1,p+1)=1$ that $b(0,v)=0$
whenever $v\ge 1$. In case $q=v$, we have
\begin{eqnarray*}
b(q,q)\mbox{\hspace{-8pt}}&=&\mbox{\hspace{-8pt}}\sum_{p=0}^q (-1)^p\left({q \atop p}\right) s(p+1,p-q+1)\\
      \mbox{\hspace{-8pt}}&=&\mbox{\hspace{-8pt}}(-1)^q s(q+1,1)=q!
\end{eqnarray*}
using (\ref{Eq_s2}) and since $s(p+1,p-q+1)=0$ for $p-q+1\le 0$.
Next, we prove the recursion for $q\ge 1$ and $v\ge 2$.
Using the recurrence formula for the binomial coefficients followed by the
recurrence formula (\ref{Eq_Rs}) for Stirling numbers of the first kind,
we obtain
\begin{eqnarray}
b(q,v)\mbox{\hspace{-8pt}}&=&\mbox{\hspace{-8pt}}\sum_{p=0}^v (-1)^p\left(\left({v-1\atop p}\right)+
             \left({v-1\atop p-1}\right)\right) \nonumber \\
      & &\mbox{\hspace{1pt}}\cdot s(p+1,p-q+1)  \nonumber \\
      \mbox{\hspace{-8pt}}&=&\mbox{\hspace{-8pt}}\sum_{p=0}^{v-1} (-1)^p\left({v-1\atop p}\right)s(p+1,p-q+1)\nonumber \\
      & &\mbox{\hspace{1pt}}+\sum_{p=0}^v(-1)^p\left({v-1\atop p-1}\right)\nonumber \\
      & &\mbox{\hspace{1pt}}\cdot \left(s(p,p-q)-ps(p,p-q+1)\right)\nonumber \\
      \mbox{\hspace{-8pt}}&=&\mbox{\hspace{-8pt}}-\sum_{p=0}^v(-1)^p\left({v-1\atop p-1}\right)ps(p,p-q+1) \label{Eq_rec}\\
      \mbox{\hspace{-8pt}}&=&\mbox{\hspace{-8pt}}\sum_{p=0}^{v-1}(-1)^p\left({v-1\atop p}\right)(p+1)\nonumber \\
      & &\mbox{\hspace{1pt}}\cdot s(p+1,p-q+2)\nonumber \\
      \mbox{\hspace{-8pt}}&=&\mbox{\hspace{-8pt}}\sum_{p=0}^{v-1}(-1)^p \left({v-1\atop p}\right)s(p+1,p-q+2)\nonumber \\
      & &\mbox{\hspace{-5pt}}+\sum_{p=0}^{v-1}(-1)^p\left({v-1\atop p}\right)ps(p+1,p-q+2)\nonumber.
\end{eqnarray}
Next, we use
$$\left({v-1\atop p}\right)p=\left({v-2\atop p-1}\right)(v-1)$$
followed by (\ref{Eq_Rs}) to conclude that
\begin{eqnarray*}
b(q,v)\mbox{\hspace{-8pt}}&=&\mbox{\hspace{-8pt}}b(q-1,v-1)+(v-1)\sum_{p=0}^{v-1}(-1)^p \left({v-2\atop p-1}\right)\\
      & &\cdot s(p+1,p-q+2)\\
      \mbox{\hspace{-8pt}}&=&\mbox{\hspace{-8pt}}b(q-1,v-1)+(v-1)\sum_{p=0}^{v-1}(-1)^p \left({v-2\atop p-1}\right)\\
      & &\cdot (s(p,p-q+1)-ps(p,p-q+2))\\
      \mbox{\hspace{-8pt}}&=&\mbox{\hspace{-8pt}}b(q-1,v-1)\\
      & &\mbox{\hspace{1pt}}-(v-1)\sum_{p=0}^{v-2}(-1)^p \left({v-2\atop p}\right)s(p+1,p-q+2)\\
      & &\mbox{\hspace{1pt}}-(v-1)\sum_{p=0}^{v-1}(-1)^p \left({v-2\atop p-1}\right)p s(p,p-q+2)\\
      \mbox{\hspace{-8pt}}&=&\mbox{\hspace{-8pt}}b(q-1,v-1)-(v-1)b(q-1,v-2)\\
      & &\mbox{\hspace{1pt}}+(v-1)b(q-1,v-1),
\end{eqnarray*}
where we used (\ref{Eq_b}) and (\ref{Eq_rec}) after replacing $v$ by $v-1$ and $q$ by $q-1$.
Hence, we have
$$b(q,v)=vb(q-1,v-1)-(v-1)b(q-1,v-2).$$
\endofproof

The values of $b(q,v)$ for $0\le q,v\le 7$ are listed in Table I.
\begin{table}
\caption{$b(q,v)$ for $0\le q,v\le 7$}
$$\begin{array}{|c||c|c|c|c|c|c|c|c|} \hline
q\backslash v& 0 & 1 & 2 & 3 & 4 & 5 & 6 & 7  \\ \hline\hline
0            & 1 & 0 & 0 & 0 & 0 & 0 & 0 & 0  \\ \hline
1            & 0 & 1 &-1 & 0 & 0 & 0 & 0 & 0  \\ \hline
2            & 0 & 0 & 2 &-5 & 3 & 0 & 0 & 0  \\ \hline
3            & 0 & 0 & 0 & 6 &-26&35&-15&  0  \\ \hline
4            & 0 & 0 & 0 & 0 &24 &-154& 340&-315 \\ \hline
5            & 0 & 0 & 0 & 0 & 0 & 120 & -1044 & 3304 \\ \hline
6            & 0 & 0 & 0 & 0 & 0 & 0 & 720 & -8028 \\ \hline
7            & 0 & 0 & 0 & 0 & 0 & 0 &   0 & 5040 \\ \hline
\end{array}$$
\end{table}

Now we state an explicit expression for $b(q,v)$. For this purpose, we
write $a<_2 b$ for real numbers $a$ and $b$ if and only if $b-a\ge 2$.
For example, $3<_2 5$ but $3\not<_2 4$.

\begin{lemma} \label{Lem_Ham5}
For nonnegative integers $q$ and $v\ge q$, we have
\begin{eqnarray}
b(q,v)\hspace{-8pt}&=&\hspace{-8pt}(-1)^{v-q} v! \sum_{0=k_0<_2 k_1<_2\cdots<_2 k_{v-q}<_2 k_{v-q+1}=v+2}\nonumber\\
          & &  \prod_{i=1}^{v-q}{1\over k_i}.  \label{Eq_Ham51}
\end{eqnarray}
\end{lemma}
\beginofproof Let $\bar{b}(q,v)$ denote the expression in the right hand side of (\ref{Eq_Ham51}).
It suffices to verify that $\bar{b}(q,v)$ satisfies the recursion given in Lemma~\ref{Lem_Ham4}
for $q\ge 1$ and $v\ge 2$ and that $\bar{b}(q,v)$ and $b(q,v)$ agree in the cases $q=0<v$ and $q=v$.
First, we notice that in case $0=q<v$, the summation in the expression of $\bar{b}(q,v)$ is over all tuples
$0=k_0,k_1,\ldots,k_v,k_{v+1}=v+2$ for which $k_i-k_{i-1}\ge 2$ for $i=1,2,\ldots,v+1$. For such tuples
\vspace{-4pt}
$$v+2=k_{v+1}-k_0=\sum_{i=1}^{v+1}(k_i-k_{i-1})\ge 2(v+1)> v+2.$$
\hspace{-4pt}
This contradiction shows that the sum in the expression of
$\bar{b}(q,v)$ is empty and, therefore, $\bar{b}(0,v)=0$ for $v\ge 1$, which is
identical to $b(0,v)$ given in Lemma~\ref{Lem_Ham4}.
Next, we notice that in case $v=q$, the summation in the expression of  $\bar{b}(q,v)$ is over one
tuple only, $(k_0,k_1)$ with $k_0=0$ and $k_1=q+2$, while the product is empty, and hence equals one. This
verifies that $\bar{b}(q,q)=q!$, which is also
identical to $b(q,q)$ given in Lemma~\ref{Lem_Ham4}.
Finally, we verify the recursion
\begin{equation}  \label{Eq_Ham511}
\bar{b}(q,v)=v\bar{b}(q-1,v-1)-(v-1)\bar{b}(q-1,v-2)
\end{equation}
for $q\ge 1$ and $v\ge 2$. We notice that
\begin{eqnarray}
v\bar{b}(q-1,v-1)\hspace{-8pt}&=&\hspace{-8pt}v(-1)^{v-q} (v-1)! \nonumber \\
          & &\hspace{-60pt} \sum_{0=k_0<_2 k_1<_2\cdots<_2 k_{v-q}<_2 k_{v-q+1}=v+1}\hspace{10pt}
            \prod_{i=1}^{v-q}{1\over k_i} \nonumber \\
                 \hspace{-8pt}&=&\hspace{-8pt}(-1)^{v-q} v! \nonumber \\
          & &\hspace{-60pt} \sum_{0=k_0<_2 k_1<_2\cdots<_2 k_{v-q}<_2 k_{v-q+1}=v+1}\hspace{10pt}
            \prod_{i=1}^{v-q}{1\over k_i} \label{Eq_Ham52}
\end{eqnarray}
and that
\begin{eqnarray}
(v-1)\bar{b}(q-1,v-2)\hspace{-8pt}&=&\hspace{-8pt}(v-1)(-1)^{v-q-1} (v-2)! \nonumber \\
                   & &\hspace{-60pt} \sum_{0=k_0<_2 k_1<_2\cdots<_2 k_{v-q}=v}
            \prod_{i=1}^{v-q-1}{1\over k_i} \nonumber \\
                 \hspace{-8pt}&=&\hspace{-8pt}-(-1)^{v-q} v! \nonumber \\
                  & &\hspace{-60pt}  \sum_{0=k_0<_2 k_1<_2\cdots<_2 k_{v-q}=v}\hspace{10pt}
            \prod_{i=1}^{v-q-1}{1\over k_iv}. \label{Eq_Ham53}
\end{eqnarray}
We notice that the sets of tuples $k_0,\ldots,k_{v-q}$ over which the summations in (\ref{Eq_Ham52}) and (\ref{Eq_Ham53}) run
are disjoint and their union gives the set of tuples over which the summation in the expression of $\bar{b}(q,v)$ runs.
Hence, (\ref{Eq_Ham511}) holds.
\endofproof

The following corollary gives values of $b(q,v)$ for selected pairs $(q,v)$.
\begin{corollary} \label{Cor1}
For nonnegative integers $q$,
\begin{eqnarray*}
b(q,q)&=&q! \\
b(q,q+1)&=&-(q+1)!\sum_{k=2}^{q+1} {1\over k}\\
b(q,2q)&=&(-1)^q\prod_{i=0}^{q-1}(2q-2i-1)\\
b(q,v)&=&0 \mbox{ \ \ for $v>2q$.}
\end{eqnarray*}
\end{corollary}
\beginofproof
The statement regarding $b(q,q)$ follows from Lemma~\ref{Lem_Ham4}. The expression for $b(q,q+1)$ follows readily from
(\ref{Eq_Ham51}). For $v\ge 2q$, we notice that the summation in the right hand side of (\ref{Eq_Ham51}) runs over
all tuples $0=k_0,k_1,\ldots,k_{v-q},k_{v-q+1}=v+2$ for which $k_i-k_{i-1}\ge 2$ for $i=1,2,\ldots,v-q+1$.
For such tuples,
\hspace{-4pt}
$$v+2=k_{v-q+1}-k_0=\sum_{i=1}^{v-q+1}(k_i-k_{i-1})\ge 2(v-q+1).$$
\hspace{-4pt}
Hence, for $v>2q$, the sum in the expression of
$b(q,v)$is empty and $b(q,2v)=0$. For $v=2q$, the summation is over one tuple only for which
$k_i=2i$ for $i=0,1,\ldots,q+1$. This proves the expression of $b(q,2q)$.
\endofproof

Combining Lemmas~\ref{Lem_Ham3}, \ref{Lem_Ham4}, and \ref{Lem_Ham5}, we obtain the major result of this correspondence.
\begin{theorem} \label{Thm2}
Let $\bf H$ be an $m\times (2^m-1)$ parity-check matrix of a Hamming code of
length $n=2^m-1$.
Then, for $l=0,1,\ldots,2^m-1$,
$$S_l={1\over l!}\sum_{q=0}^l \sum_{v=q}^{\min\{2q,l\}} (-1)^v\left({l\atop v}\right)b(q,v)(2^{l-q}-(l-v))^m$$
where $b(q,v)$ is given in Lemmas~\ref{Lem_Ham4} and \ref{Lem_Ham5}.
\end{theorem}

\begin{corollary} \label{Cor2}
Let $\bf H$ be an $m\times (2^m-1)$ parity-check matrix of a Hamming code of
length $n=2^m-1$.
Then, for a fixed value of
$l\ge 3$, $S_l$ behaves asymptotically as a function of the code length $n$ as
\begin{equation} \label{Slasym}
S_l \sim {n^{\log_2(2^l-l)}\over l!}.
\end{equation}
\end{corollary}

It is interesting to compare the expression of $S_l$ with that of $A_l$,
the number of codewords of weight $l$ in a Hamming code of length $n=2^m-1$,
given by
$$A_l={1\over n+1}\left[\left({n \atop l}\right)+(-1)^{\lceil l/2\rceil}n\left({(n-1)/2\atop \lfloor l/2\rfloor}\right)\right].$$
This expression can be verified by noticing that $A_0=1$, $A_1=0$,
and that it satisifies the recursion stated on page 129 of \cite{MS}.
For a fixed value of $l\ge 3$, $A_l$ behaves asymptotically as a function of the code length $n$ as
\begin{equation} \label{Alasym}
A_l \sim {n^{l-1}\over l!}.
\end{equation}
                                                                                                                                                             
The asymptotic expressions of $S_l$ and $A_l$ for $l\ge 3$ can be directly justified as follows.
First notice that $l! S_l$ is the number of $m\times l$ binary matrices with
different non-zero columns such that no row has weight one.
The total number of $m\times l$ binary matrices without any row of weight one is
$(2^l-l)^m$. Amongst these matrices, there are at most $l2^{(l-1)m}$ matrices with at least one
all-zero column and at most $({l \atop 2})2^{(l-1)m}$ matrices with at least two identical columns
for $l\ge 2$.
Therefore, by the union bound we get
$$(2^l-l)^m-\left[l2^{(l-1)m}+\left(l\atop 2\right)2^{(l-1)m}\right]\le l! S_l
  \le (2^l-l)^m.$$
Hence, for a fixed $l\ge 3$, as $n=2^m-1$ tends to infinity, the ratio $l! S_l/(2^l-l)^m$ tends to one, and (\ref{Slasym})
follows.
                                                                                                                                                             
A similar argument leads to (\ref{Alasym}). Notice that
$l! A_l$ is the number of $m\times l$ matrices with different non-zero columns and
even-weight rows. The total number of $m\times l$ binary matrices with even-weight rows is
$2^{(l-1)m}$. Amongst these matrices, there are at most $l2^{(l-2)m}$ matrices with at least one
all-zero column and at most $({l \atop 2})2^{(l-2)m}$ matrices with at least two identical columns
for $l\ge 3$.
Therefore, by the union bound we get
$$2^{(l-1)m}-\left[l2^{(l-2)m}+\left(l\atop 2\right)2^{(l-2)m}\right]\le l! A_l
  \le 2^{(l-1)m}.$$
Hence, for a fixed $l\ge 3$, as $n=2^m-1$ tends to infinity, the ratio $l! A_l/2^{(l-1)m}$ tends to one, and (\ref{Alasym})
follows.

Direct application of Theorem~\ref{Thm2} gives the following results including
the expression of $S_3$ in (\ref{eqME}) stated by McEliece \cite{McE}
\begin{eqnarray*}
S_0\hspace{-8pt}&=&\hspace{-8pt}1, S_1=S_2=0,\\
S_3\hspace{-8pt}&=&\hspace{-8pt}{1 \over 6}\left(5^m-3\times 3^m+2\times 2^m\right),\\
S_4\hspace{-8pt}&=&\hspace{-8pt}{1\over 24}\left(12^m-6\times 6^m-4\times 5^m+3\times 4^m \right.\\
                & &\left. +20\times 3^m-14\times 2^m\right),\\
S_5\hspace{-8pt}&=&\hspace{-8pt}{1\over 120}\left(27^m-10\times 13^m-5\times 12^m\right.\\
                & &+15\times 7^m+50\times 6^m+20\times 5^m\\
                & &\left. -35\times 4^m-130\times 3^m+94\times 2^m \right).\\
\end{eqnarray*}

Furthermore, if $S_m(x)$ denotes the stopping set enumerator for the full-rank parity-check
matrix of a Hamming code of length $2^m-1$, then
\begin{eqnarray*}
S_3(x)\hspace{-8pt}&=&\hspace{-8pt}1+10 x^3+23 x^4+21 x^5+7 x^6+x^7,\\
S_4(x)\hspace{-8pt}&=&\hspace{-8pt}1+69 x^3+526 x^4+1979 x^5+ 4333 x^6+ 6211 x^7\\
                   & &\hspace{-2pt}           + 6403 x^8 +5005 x^9+ 3003 x^{10}+1365 x^{11}\\
                   & &\hspace{-2pt}            + 455 x^{12}+105 x^{13}+15 x^{14}+x^{15},\\
S_5(x)\hspace{-8pt}&=&\hspace{-8pt}1+410 x^3+8215 x^4+83590 x^5+519481 x^6\\
                   & &\hspace{-2pt}            +2243175 x^7+7378485 x^8+19645915 x^9\\
                   & &\hspace{-2pt}            +43951765 x^{10}+84432075 x^{11}+141011325 x^{12}\\
                   & &\hspace{-2pt}            +206216675 x^{13}+265174125 x^{14}+300538995 x^{15}\\
                   & &\hspace{-2pt}            +300540115 x^{16}+265182525 x^{17}+206253075 x^{18}\\
                   & &\hspace{-2pt}            +141120525 x^{19}+84672315 x^{20}+44352165 x^{21}\\
                   & &\hspace{-2pt}            +20160075 x^{22} +7888725  x^{23}+ 2629575 x^{24}\\
                   & &\hspace{-2pt}            +736281 x^{25}+169911 x^{26}+31465 x^{27}\\
                   & &\hspace{-2pt}            +4495 x^{28}+465 x^{29}+ 31 x^{30} + x^{31}.
\end{eqnarray*}
\begin{center}
{\bf Acknowledgment}
\end{center}
The authors would like to thank Dr. Ludo Tolhuizen for valuable comments.

\end{document}